\documentclass[twocolumn,italian,english,english]{revtex4}
\usepackage[T1]{fontenc}
\usepackage[latin1]{inputenc}
\usepackage{color}
\usepackage{graphicx}

\makeatletter

\providecommand{\LyX}{L\kern-.1667em\lower.25em\hbox{Y}\kern-.125emX\@}
\makeatletter

\usepackage{babel}
\makeatother
\begin{document}

\title{Supernovae type Ia data favour coupled phantom energy}

\author{Elisabetta Majerotto }

\email{majerotto@mporzio.astro.it, }

\author{Domenico Sapone }

\email{sapone@mporzio.astro.it}

\author{Luca Amendola}

\email{amendola@mporzio.astro.it}

\affiliation{INAF/Osservatorio Astronomico di Roma\\
Via Frascati 33, 00040 Monteporzio Catone (Roma), Italia}

\begin{abstract}
We estimate the constraints that the recent high-redshift sample of
supernovae type Ia put on a phenomenological interaction between dark
energy and dark matter. The interaction can be interpreted as arising
from the time variation of the mass of dark matter particles. We find
that the coupling correlates with the equation of state: roughly speaking,
a negative coupling (in our sign convention) implies phantom energy
($w_{\phi }<-1$) while a positive coupling implies {}``ordinary''
dark energy. The constraints from the current supernovae Ia Hubble
diagram favour a negative coupling and an equation of state $w_{\phi }<-1$.
A zero or positive coupling is in fact unlikely at 99\% c.l. (assuming
constant equation of state); at the same time non-phantom values ($w_{\phi }>-1$)
are unlikely at 95\%. We show also that the usual bounds on the energy
density weaken considerably when the coupling is introduced: values
as large as $\Omega _{m0}=0.7$ become acceptable for as concerns
SNIa. We find that the rate of change of the mass $\dot{m}/m$ of
the dark matter particles is constrained to be $\delta _{0}$ in a
Hubble time, with $-10<\delta _{0}<-1$ to 95\% c.l.. We show that
a large positive coupling might in principle avoid the future singularity
known as {}``big rip'' (occurring for $w_{\phi }<-1$) but the parameter
region for this to occur is almost excluded by the data. We also forecast
the constraints that can be obtained from future experiments, focusing
on supernovae and baryon oscillations in the power spectra of deep
redshift surveys. We show that the method of baryon oscillations holds
the best potential to contrain the coupling.
\end{abstract}
\maketitle

\section{Introduction}

One of the main open problems in Cosmology is to determine the properties
of the unclustered component called dark energy that is required to
explain CMB, supernovae Ia, cluster masses and other observational
data.

Although most work on dark energy assumes it to be coupled to the
other field only through gravity, there is by now a rich literature
on possible interactions to standard fields and to dark matter. Even
before the evidences in favor of dark energy, a coupling between a
cosmic scalar field and ordinary matter has been studied by Wetterich
\cite{wet88,wet90} while Damour, Gundlach and Gibbons \cite{dam}
investigated the possibility of a species-dependent coupling within
the context of Brans-Dicke models. Generally speaking, any low-energy
limit of higher-dimensional theories predicts the existence of scalar
fields coupled to matter. More recently, several authors considered
an explicit coupling between dark energy and matter: we denote this
class of models as \emph{coupled dark energy}. A partial list of works
in this field is in refs. \cite{cq,chim,post2002}. Other authors
considered a coupling to specific standard model fields: to the electromagnetic
field \cite{carroll}, to neutrinos \cite{horvat}, to baryonic or
leptonic current \cite{li}: in these cases, there are strong constraints
from local observations or from variation of fundamental constants. 

In this paper we confine our attention to the interaction with dark
matter, as in ref. \cite{cq}. Such a coupling is of course observable
only with astrophysical \cite{gf} and cosmological \cite{aq} experiments
involving growth of perturbations or global geometric effects. Notwithstanding
the large number of papers dealing with variants of coupled dark energy,
the works dedicated to constraining the interaction via supernovae
Type Ia (SNIa) are very limited. The main reason is that there is
yet no compelling form of the interaction (just as there is no compelling
form of the dark energy equation of state). To make a step towards
constraining the energy exchange between the dark components, instead
of considering a specific coupling motivated by (or inspired by) some
fundamental theory, we adopt here a phenomenological point of view.
That is, we assume a general relation between dark energy and dark
matter and derive its theoretical and observational properties. Our
simple relation includes several previous coupled and uncoupled dark
energy models but extends the analysis to cases which have not been
tested so far. The main aim of this paper is to derive bounds on the
interaction strength by analysing the recent SNIa sample of Riess
et al. \cite{riess2004}, which include several $z>1$ supernovae.
This works extends previous investigations in ref. \cite{Dalal},
which studied the same relation between dark matter and dark energy,
and in ref. \cite{agp}, which compared the recent SNIa data with
a special class of coupled dark energy models motivated by superstrings
and characterized by a constant ratio of densities.

\section{Modeling the interaction}

Let us start with a model containing only dark matter and dark energy.
The basic assumption of this paper is that the dark components interact
through an energy exchange term. The conservation equations in FRW
metric with scale factor $a$ can be written in all generality as\begin{equation}
\begin{array}{ccc}
 \dot{\rho }_{m}+3H\rho _{m} & = & \delta H\rho _{m}\, ,\\
 \dot{\rho }_{_{\phi }}+3H\rho _{\phi }(1+\omega _{\phi }) & = & -\delta H\rho _{m}\, ,\end{array}\label{contcoup}\end{equation}
where the subscript $m$ stands for dark matter while the subscript
$\phi $ stands for dark energy and where $\delta $ is a dimensionless
coupling function. In principle the coupling $\delta $ might depend
on all degrees of freedom of the two components. However, if $\delta $
is a function of the scale factor $a$ only then the first equation
of (\ref{contcoup}) can be integrated out to give (we put the present
value $a_{0}=1$)\begin{equation}
\rho _{m}=\rho _{m0}a^{-3}e^{\int \delta d\alpha }\, ,\label{eq:matter}\end{equation}
where $\alpha =\log a$. This shows that the interaction causes $\rho _{m}$
to deviate from the standard scaling $a^{-3}$. That is, matter is
not {}``conserved'' or, equivalently, the mass $m$ of matter particles
$\rho _{m}=m(a)n$, where $n$ is their number density, varies with
time such that\begin{equation}
\frac{m'}{m}=\delta \label{eq:vamps}\end{equation}
where the prime denotes derivation with respect to $\alpha $. Therefore,
$\delta $ can be interpreted as the rate of change of the particle
mass per Hubble time.

While so far many paper have been devoted to find constraints on $w_{\phi }(a)$,
aim of this paper is to derive constraints on $\delta (a)$ from SNIa.
We need therefore a sensible parametrization of $\delta (a)$. Generally
speaking, we are faced with two possibilities: either we write down
a parametrization for $\delta (a)$ and then find from this the relation
between $\rho _{m}$ and $\rho _{\phi }$, or first give the latter
and derive a function $\delta (a)$. We find that this second choice
is simpler and better connected to previous work and constraints. 

The basic relation we start from is that the fluid densities scale
according to the following relation \cite{Dalal}\begin{equation}
\rho _{\phi }/\rho _{m}=Aa^{\xi }\label{rap}\end{equation}
 where $A,\xi $ are two constant parameters. Moreover, we approximate
$w_{\phi }$ as a constant: it is clear however that a more complete
analysis should allow for a time-dependent equation of state. The
relation (\ref{rap}) has two useful properties: \emph{a}) it includes
all scaling solutions (defined as those with $\rho \sim a^{m}$) and
\emph{b}) the functions $\rho _{m}(a),\rho _{\phi }(a)$ can be calculated
analytically. Since for uncoupled dark energy models with constant
equation of state one has $\rho _{m}\sim a^{-3}$ and $\rho _{\phi }\sim a^{-3(1+w_{\phi })}$
it appears that the relation (\ref{rap}) reduces to this case for
$\xi =-3w_{\phi }$. Conversely, if $\xi \not =-3w_{\phi }$ then
the matter density deviates from the $a^{-3}$ law, as we have seen
above.

The relation (\ref{rap}) has been first introduced and tested in
ref. \cite{Dalal}. The present paper extends their investigation
in several respects. First, we include baryons. Ref. \cite{Dalal}
assumed a single matter component, but there are strong upper limits
to an interaction of dark energy with baryons (i.e. upper limits to
a non-conservation relation like Eq. (\ref{eq:matter}) for baryons,
see e.g. \cite{hagi}). We will assume therefore that baryons are
uncoupled while the coupling to dark matter is left as a free parameter.
This has important consequences for the general behavior of the model.
Second, ref. \cite{Dalal} confined the range of parameters to $0<\xi <3$
and $w_{\phi }>-1$, while we extend the range to a much larger domain,
thereby including also the regime of {}``phantom'' dark energy \cite{cald}.
Third, we use the new data of Riess et al. \cite{riess2004}: these
include SN at $z>1$ and extend therefore by a considerable factor
the leverage arm of the method. Fourth, we'll discuss the implication
of the results for as concerns the beginning of the acceleration.
In \cite{riess2004} it was claimed that the new high-redshift SN
constrain $z_{acc}<1$ (more exactly, $z_{acc}=0.46\pm 0.13$). In
contrast with this, we show that $z_{acc}>1$ is not ruled out at
more than 95\% when the coupling is non-zero. A similar conclusion
is obtained in ref. \cite{agp} studying a string-motivated form of
interaction and in \cite{bck} adopting different parametrizations
of the equation of state. Finally, we will produce forecasts for future
experiments. 

Before we include the baryons, let us derive some relations in the
simplified case of two coupled components only. Assuming $\Omega _{tot}=\Omega _{\phi }+\Omega _{m}=1$
we obtain from (\ref{rap}):\begin{equation}
\Omega _{\phi }+\frac{1}{A}a^{-\xi }\Omega _{\phi }=1\, ,\label{flat}\end{equation}
 so that the constant $A$ can be expressed in function of $\Omega _{\phi ,0}$
(the subscript $0$ indicates the present epoch)

\[
A=\frac{\Omega _{\phi ,0}}{1-\Omega _{\phi ,0}}\, .\]

If the dark matter is pressureless, $w_{m}=0$, the total energy density
obeys the conservation equation\begin{equation}
\frac{d\rho _{tot}}{da}+\frac{3}{a}(1+\omega _{\phi }\Omega _{\phi })\rho _{tot}=0\, .\label{eqcont}\end{equation}
 Therefore, assuming a constant $w_{\phi }$, we obtain\begin{equation}
\rho _{tot}=\rho _{0}a^{-3}[1-\Omega _{\phi ,0}(1-a^{\xi })]^{-3\frac{\omega _{\phi }}{\xi }}\, ,\label{rhotot}\end{equation}
 and the Friedman equation\begin{equation}
H^{2}=H_{0}^{2}a^{-3}[1-\Omega _{\phi ,0}(1-a^{\xi })]^{-3\frac{\omega _{\phi }}{\xi }}\, .\label{eq:fri1}\end{equation}
It is worth remarking that this expression for $H^{2}$ cannot be
reproduced by a simple model of varying $w$, as e.g. $w(z)=w_{0}+w_{1}z$.
This means that the effect of the coupling is intrinsically different
from the effect of a time-dependent equation of state.

The relation between the scaling behavior and the coupling $\delta $
is easily derived by imposing the relation (\ref{rap})\begin{equation}
\frac{d}{dt}(\frac{\rho _{\phi }}{\rho _{m}a^{\xi }})=0\, .\label{staz}\end{equation}
 We find then\begin{equation}
\delta =-\frac{(\xi +3\omega _{\phi })}{\rho _{\phi }+\rho _{m}}\rho _{\phi }\, .\label{del1}\end{equation}

Substituting (\ref{rap}) we obtain the evolution of the coupling
as a function of $a$\begin{equation}
\delta =\frac{\delta _{0}}{\Omega _{\phi ,0}+(1-\Omega _{\phi ,0})a^{-\xi }}\, ,\label{deltaa}\end{equation}
where\begin{equation}
\delta _{0}=-\Omega _{\phi ,0}(\xi +3\omega _{\phi })\, .\label{delta0}\end{equation}

\section{Adding the baryons}

So far we assumed a single matter component. However, as already anticipated,
if dark energy interacts with baryons then a new long-range force
arises, on which there are strong experimental upper limits \cite{hagi,dam}.
The baryonic component has therefore to be assumed extremely weakly
coupled; for simplicity, we assume here that the baryons are totally
uncoupled. In principle, of course, we should allow the possibility
that also part of the dark matter itself is uncoupled, as in Ref.
\cite{agp} but for simplicity we restrict ourselves to the basic
case in which all dark matter couples with the same strength. Here
we derive the corresponding formulae when uncoupled baryons are added
to the cosmic fluid. As it will appear clear, adding a small percentage
of baryons at the present does not change qualitatively the fit to
the supernovae; however, it changes dramatically the past and future
asymptotic behavior of the cosmological model. In all the plots and
numerical results we always assume $\Omega _{b,0}=0.05$. The parameter
$A$ now becomes

\[
A=\frac{\Omega _{\phi ,0}}{1-\Omega _{b,0}-\Omega _{\phi ,0}}\, ,\]
 from which\begin{equation}
\delta =\frac{\delta _{0}(1-\Omega _{b,0})}{\Omega _{\phi ,0}+(1-\Omega _{b,0}-\Omega _{\phi ,0})a^{-\xi }}\, ,\label{deltaa}\end{equation}
so that the present coupling is

\begin{equation}
\delta _{0}=-\frac{\Omega _{\phi ,0}}{1-\Omega _{b,0}}(\xi +3\omega _{\phi })\, .\label{deltabar}\end{equation}
As it appears from Fig. 1, when $\xi >0$ the coupling varies from
zero in the past to a constant value $\delta _{1}=-(\xi +3w_{\phi })$
in the future. As expected, the standard model of uncoupled perfect
fluid dark energy is recovered when $\xi =-3w_{\phi }$. For $\xi <0$
the behavior is opposite. On the other hand, models with a constant
$\delta $ (investigated in \cite{agp}) are obtained for $\xi =0$.
In this case, $\rho _{\phi }\sim \rho _{m}$: such behavior has been
called {}``tracking'' dark energy when the regime is transient \cite{cald},
and {}``stationary dark energy'' \cite{cq,bias} when the regime
is the final attractor. To emphasize the connection between $\xi $
and the coupling, we will use in most cases the present coupling $\delta _{0}=\delta (a=1)$
as free parameter instead of $\xi $. The parameters that characterize
the model are therefore $(\Omega _{\phi ,0},\omega _{\phi },\delta _{0})$.
$\Lambda $CDM corresponds to $w_{\phi }=-1$ and $\delta _{0}=0$.
The trend of the coupling $\delta $ can be expanded for low redshifts
as \textcolor{red}{\[
\delta \simeq \delta _{0}(1-z\frac{\xi \Omega _{m,0}}{1-\Omega _{b,0}})\]
}The sign of $\xi $ is therefore also associated to the sign of the
time derivative of the coupling. 

The Friedman equation is now\begin{equation}
H^{2}=H_{0}^{2}E(a)^{2}\, ,\label{friedbar}\end{equation}
 where\begin{equation}
E^{2}=(1-\Omega _{b,0})a^{-3}[1-\frac{\Omega _{\phi ,0}}{1-\Omega _{b,0}}(1-a^{\xi })]^{-3\frac{\omega _{\phi }}{\xi }}+\Omega _{b,0}a^{-3}\, ,\label{friedbare}\end{equation}
 which for $\xi =0$ becomes\begin{equation}
H^{2}=H_{0}^{2}\{(1-\Omega _{b,0})a^{-\beta }+\Omega _{b,0}a^{-3}\}\, ,\label{friedbarx0}\end{equation}
 where $\beta =\frac{1-\Omega _{b,0}+3\omega _{\phi }\Omega _{\phi ,0}}{1-\Omega _{b,0}}$.

It is interesting to derive the asymptotic limits of $\Omega _{b,m,\phi }$
(Figs. 2-4) Notice that, assuming $w_{\phi }<0$, the asymptotic future
and past behaviour of the total density and of the density parameters
of all the components depends only on the parameters $\xi $ . For
$\xi >0$ we have the limits for $a\rightarrow 0$

\begin{eqnarray}
\Omega _{b} & \rightarrow  & \frac{\Omega _{b,0}}{(1-\Omega _{b,0})[1-\frac{\Omega _{\phi ,0}}{1-\Omega _{b,0}}]^{-3\omega _{\phi }/\xi }+\Omega _{b,0}}\, ,\label{limomegab+}\\
\Omega _{m} & \rightarrow  & 1-\Omega _{b}\, ,\label{limomegam+}\\
\Omega _{\phi } & \rightarrow  & 0\, ,\label{limomegaphi+}
\end{eqnarray}
 while for $\xi \le 0$ we have $\Omega _{b}\to 1$ and $\Omega _{m},\Omega _{\phi }\to 0$,
so that the baryons dominated the past evolution even if their present
density is very low. The future asymptotics for $\xi >0$ is instead
always dominated by dark energy,\begin{eqnarray}
\Omega _{b} & \rightarrow  & 0\, ,\label{obinf+}\\
\Omega _{m} & \rightarrow  & 0\, ,\label{ominf+}\\
\Omega _{\phi } & \rightarrow  & 1\, ,\label{ophiinf+}
\end{eqnarray}
For $\xi <0$ we have

\begin{eqnarray}
\Omega _{b} & \rightarrow  & \frac{\Omega _{b,0}}{(1-\Omega _{b,0})[1-\frac{\Omega _{\phi ,0}}{1-\Omega _{b,0}}]^{-3\omega _{\phi }/\xi }+\Omega _{b,0}}\, ,\label{obinf-}\\
\Omega _{m} & \rightarrow  & 1-\Omega _{b}\, ,\label{ominf-}\\
\Omega _{\phi } & \rightarrow  & 0\, ,\label{ophiinf-}
\end{eqnarray}
and finally for $\xi =0$,\begin{eqnarray}
\Omega _{b} & \rightarrow  & 0\, ,\label{obinf0}\\
\Omega _{m} & \rightarrow  & \frac{\Omega _{\phi ,0}}{1-\Omega _{b,0}}\, ,\label{ominf0}\\
\Omega _{\phi } & \rightarrow  & 1-\frac{\Omega _{\phi ,0}}{1-\Omega _{b,0}}\, .\label{ophiinf0}
\end{eqnarray}

It could be assumed that the negative values of $\xi $ are to be
excluded, since they imply absence of dark matter in the past. However,
we are here studying the model only in a finite range of $z$ near
the present epoch, and for this reason even negative $\xi $ cannot
be excluded a priori. Moreover, baryons dominate in the past for $\xi <0$
and they could conceivably drive fluctuation growth. However, we will
see that in fact values $\xi <0$ are not favoured by SN data.

There are other interesting features to remark in the general behavior
of the density fractions. First, for $\delta _{0}\not =0$ the ratio
$\Omega _{b}/\Omega _{m}$ varies in time. If $\xi >0,$ the ratio
varies from a constant value in the past to infinity (if $\delta _{0}<0$)
or zero (if $\delta _{0}>0$) in the future. If $\xi \le 0$ the ratio
decreases always (for $w_{\phi }<0$): this variation in the baryon-to-matter
ratio due to the dark matter coupling could provide additional testing
ground for the coupling (e.g. \cite{barfrac}). 

Second, for $\xi <0$ the dark energy density vanishes both in the
past and in the future, for all values of the equation of state: this
means that the acceleration is only a temporary episode in the universe
history, as shown in Fig. 4. The universe was dominated by the baryons
in the past, by dark energy at the present and by a mix of baryons
and dark matter in the future (for the parameters employed in Fig.
4 the final value of $\Omega _{b}$ is very low and cannot be distinguinguished
from zero). For instance, assuming $w_{\phi }=-1,\Omega _{\phi }=0.7,\xi =-1$
it turns out that the acceleration ends in the future at $z=-0.85$.
This is a particularly striking example of how the coupling might
completely modify the past and future behavior of the cosmic evolution. \foreignlanguage{italian}{\textcolor{red}{}}

Another example comes from the existence of the future singularity
known as {}``big rip'', i.e. an infinite growth of the total energy
density in a finite time \cite{cald}. From the Friedmann equation
(\ref{friedbare}) we see that if $\xi >0$, for a large scale factor
$H^{2}$ behaves as $a^{-3(1+w_{\phi })}$, as in the standard uncoupled
model, so there is a big rip if $w_{\phi }<-1$. If, instead, $\xi <0$,
then $H^{2}\sim a^{-3}$ . This means that a negative $\xi $ prevents
the big rip for any value of $w_{\phi }$. From (\ref{deltabar})
we see that, assuming $\Omega _{b,0}\ll 1$, the big rip is prevented
when \[
\delta _{0}>3|w_{\phi }|\Omega _{\phi ,0}\]
However, we'll find that this region of parameters space is rather
disfavoured by SN data.

\begin{figure}
\begin{center}\includegraphics[  bb=50bp 500bp 598bp 843bp,
  clip,
  scale=0.5]{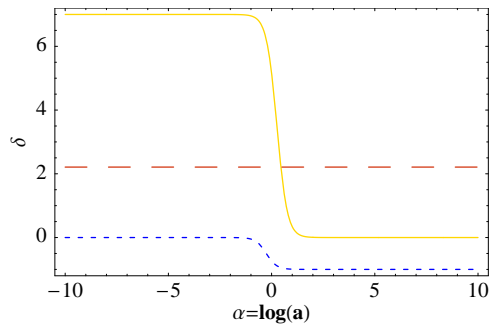}\end{center}

\caption{Behavior of $\delta (z)$ for $\Omega _{\phi }=0.7$ and $w_{\phi }=-1$.
Full line: $\xi =-4$; dotted line: $\xi =4$; dashed line: $\xi =0$.}
\end{figure}

\begin{figure}
\begin{center}\includegraphics[  clip,
  scale=0.6]{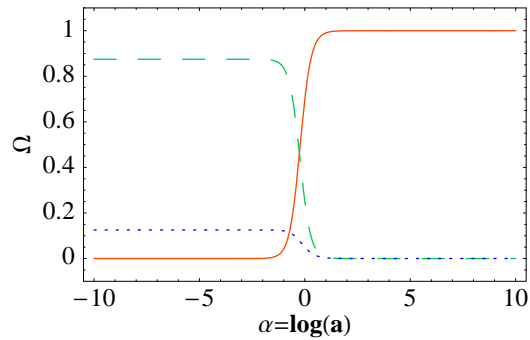}\end{center}

\caption{Behavior of $\Omega (a)$ for $\xi =4$ and $w_{\phi }=-1$, fixing
$\Omega _{b,0}=0.05,$$\Omega _{m,0}=0.25$ and $\Omega _{\phi }=0.7$.
Full line: dark energy; dotted line: baryons; dashed line: dark matter.}
\end{figure}

\begin{figure}
\begin{center}\includegraphics[  clip,
  scale=0.5]{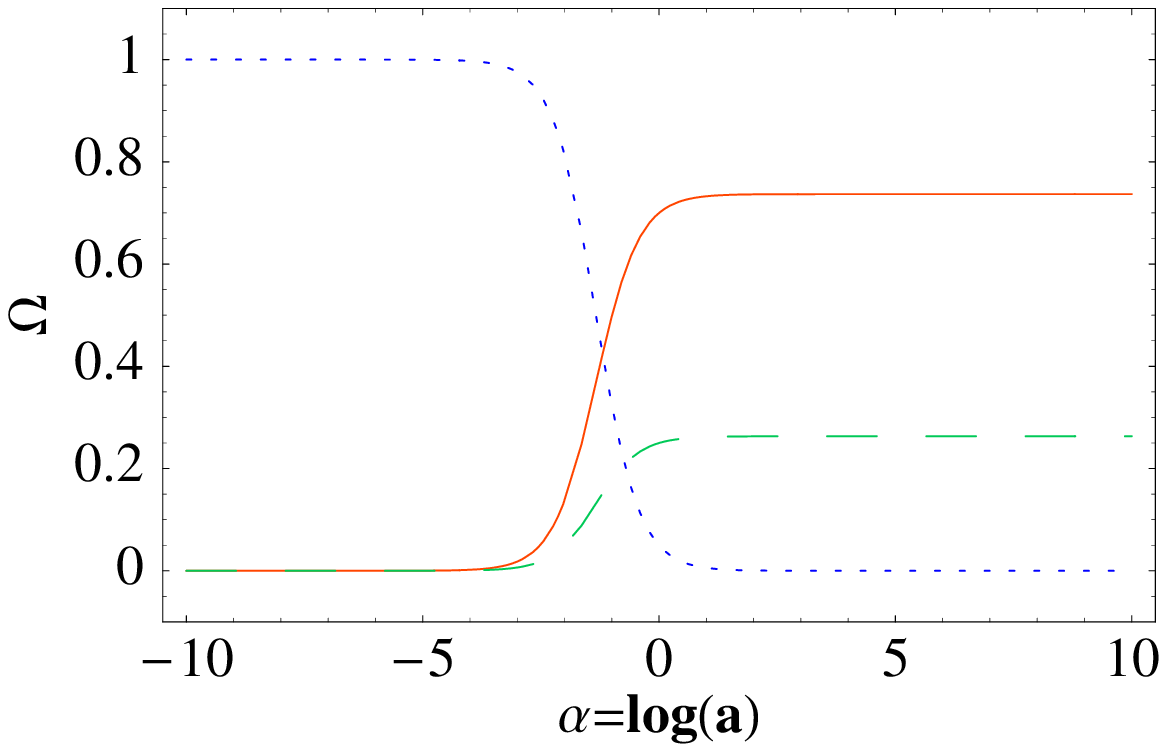}\end{center}

\caption{Behavior of $\Omega (a)$ for $\xi =0$ and $w_{\phi }=-1$ (the
other parameters are as in the previous figure). Full line: dark energy;
dotted line: baryons; dashed line: dark matter.}
\end{figure}

\begin{figure}
\begin{center}\includegraphics[  clip,
  scale=0.5]{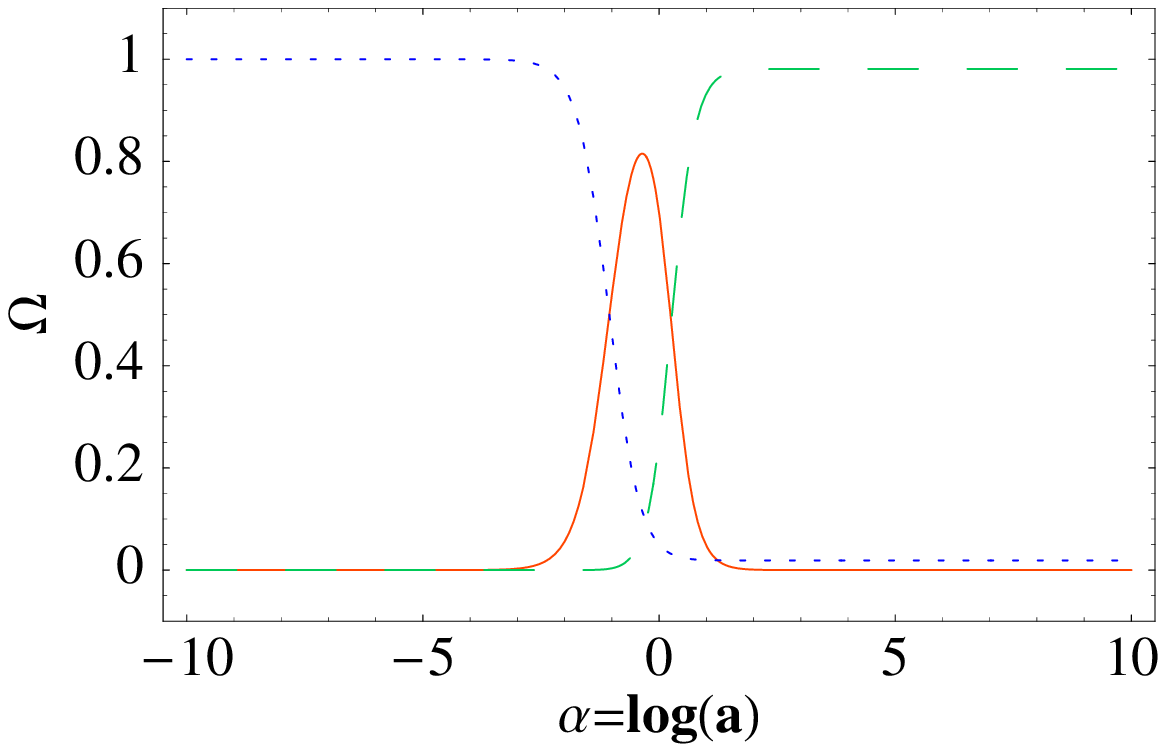}\end{center}

\caption{Behavior of $\Omega (a)$ for $\xi =-4$ and $w_{\phi }=-1$ (the
other parameters are as in the previous figure). Full line: dark energy;
dotted line: baryons; dashed line: dark matter.}
\end{figure}

\begin{figure}
\begin{center}\includegraphics[  scale=0.6]{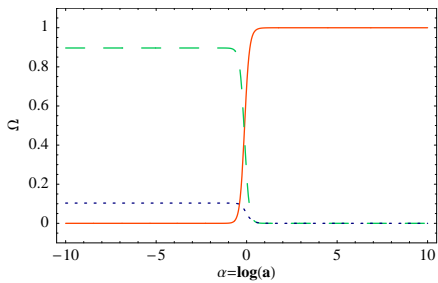}\end{center}

\caption{Behavior of $\Omega (a)$ for the best-fit values of $\xi $ and
$w_{\phi }$ (the other parameters are as in the previous figure).
Full line: dark energy; dotted line: baryons; dashed line: dark matter.}
\end{figure}

We are interested also in evaluating the epoch of acceleration. The
acceleration begins when $\ddot{a}=0$ i.e. for ( $\xi \not =0$)

\begin{eqnarray}
\frac{\ddot{a}}{a} & = & \frac{1}{2}H_{0}^{2}\{(\Omega _{b,0}-1)[1-\frac{\Omega _{\phi ,0}}{1-\Omega _{b,0}}(1-a^{\xi })]^{-3\frac{\omega _{\phi }}{\xi }}-\Omega _{b,0}\nonumber \\
 &  & -\frac{1}{2}\omega _{\phi }\Omega _{\phi ,0}a^{\xi }[1-\frac{\Omega _{\phi ,0}}{1-\Omega _{b,0}}(1-a^{\xi })]^{-3\frac{\omega _{\phi }}{\xi }-1}\}=0\nonumber \\
 &  & \label{eq:accb}
\end{eqnarray}
 which will be solved numerically later on. For $\xi =0$ the solution
is \[
z_{acc}=[\frac{1-\Omega _{b,0}-3\omega _{\phi }\Omega _{\phi ,0}}{\Omega _{b,0}}]^{\frac{1}{3-\beta }}-1\, .\]

Before starting with the data analysis, we need still another equation,
the age of the universe. This is given by\[
T=H_{0}^{-1}\int _{0}^{1}\frac{da}{aE(a)}\]
 where $E(a)$ is given in Eq. (\ref{friedbare}).

\section{Comparing with SNIa}

We used the recent compilation of SNe Ia data \cite{riess2004} (\emph{gold}
sample) to put constraints on the parameters $\Omega _{m,0},w_{\phi },\delta _{0}$
entering the expression for $H(a)$. In all subsequent plots we marginalize
over a constant offset of the apparent magnitude, so that all results
turn out to be independent of the present value of the Hubble parameter.
In all the calculations we fix $\Omega _{b,0}=0.05$.

The overall best fit is $\Omega _{\phi ,0}=0.62,w_{\phi }=-1.9,\delta _{0}=-1.5$
(see Fig. 5) with a $\chi ^{2}$=173.7 (for 157 SN), to be compared
to a $\chi ^{2}=178$ for the flat $\Lambda $CDM. In Fig. 6 we present
the main result of this paper: the contours at 68\%, 95\% and 99\%
of the likelihood function on the parameter space $\delta _{0},w_{\phi }$.
The third parameter, $\Omega _{\phi ,0}$, has been marginalized over
with a Gaussian prior $\Omega _{\phi ,0}=0.7\pm 0.1$. The $\Lambda $CDM
model lies near 1$\sigma $ from the best fit. The line $\xi =0$
is tangent to the 3$\sigma $ contour: negative values of $\xi $
lie above this line and appear to be very unlikely with respect to
positive values. It is impressive to observe how the vast majority
of the likelihood lies below the uncoupled line $\delta _{0}$, which
is almost tangential to the 1$\sigma $ contour, and leftward of $w_{\phi }=-1$.
Taken at face value, this implies that the likelihood of a negative
coupling (dark matter mass decreasing with time) contains 99\% of
the total likelihood, and that the likelihood of a phantom dark energy
($w_{\phi }<-1$) is 95\% of the total likelihood. Our conclusion
is therefore that the SNIa data are much better fitted by a negatively
coupled phantom matter than by uncoupled, positively coupled or non-phantom
stuff. It is interesting to note that it has been recently observed
\cite{amprl} that coupled phantom energy induces a repulsive interaction
(regardless of the sign of the coupling). 

In Figs. 7 and 8 we plot the 1-dimensional likelihood functions for
$\delta _{0},w_{\phi }$, marginalizing over the other parameters.
The main conclusion is that current supernovae data exclude a coupling
$\delta _{0}>0$ at 99\% c.l. if no priors are imposed on $w_{\phi }$.
The lower bound is much weaker: a negative $\delta _{0}$ down to
$\delta _{0}=-7$ is allowed to to 68\%. At 95\% level, the formal
lower limit to $\delta _{0}$ is $-10$ but the convergence to zero
is very slow and values very large and negative of $\delta _{0}$
cannot be excluded. In terms of the parameter $\xi $ we obtain $\xi >0$
at 95\% c.l. while essentially no upper limit can be put with the
same confidence. In terms of $m'/m$, this implies that the dark matter
mass can vary by a fraction $\delta _{0}$ in a Hubble time. Looking
at Fig. 6 one sees that a zero or positive coupling is instead preferred
if $w_{0}>-1$. Applying the prior $w_{\phi }>-1$ the limits on $\delta _{0}$
narrow and move to higher values: $-2.5<\delta _{0}<1.2$ (95\%).

From Fig. 8 we see that $-3<w_{\phi }<-1.2$ at 68\% and $-4.2<w_{\phi }<-1$
at 95\% c.l.. In the same Fig. 8 we plot the likelihood for $w_{\phi }$
in the uncoupled case. As it can be seen, the coupled case extends
considerably the allowed region of $w_{\phi }$, especially towards
large and negative values: a negative coupling favors phantom equation
of states. This complements the results of ref. \cite{agp} in which
it was found that a \emph{positive} coupling (corresponding to a stationary
behavior with $\xi =0$ and $w_{\phi }<0$) favours non-phantom matter.
This is a clear-cut example of how the conclusions regarding the nature
of dark energy depend crucially on its coupling to the rest of the
world. It is intriguing that the correlation between $\delta _{0}$
and $w_{\phi }$ crosses approximatively the $\Lambda $CDM case $\delta _{0}=0,w_{\phi }=-1$:
although not particularly favoured, $\Lambda $CDM remains a perfectly
acceptable model also with respect to coupled dark energy. 

We now impose on the 2D likelihood the contour levels of the acceleration
(\ref{eq:accb}). Since a negative $\delta _{0}$ means a more recent
surge of the dark energy it is to be expected that the likelihood
favours a recent acceleration. It turns out that indeed the best fit
corresponds to an acceleration epoch $z_{acc}\approx 0.3$; however,
at the same time, an acceleration $z_{acc}>1$ lies near 2$\sigma $
and cannot be excluded with large confidence (Fig. 9), especially
if one excludes phantom states of matter.

The age contours are compared to the likelihood in Fig. 10. Most of
the likelihood lies within the acceptable range 11 and 14 Gyr (for
$h=0.7$ and $\Omega _{\phi 0}=0.7$). The age constraints are therefore
rather weak.

We can also marginalize on $\delta _{0}$ and plot the likelihood
for $\Omega _{m0},w_{\phi }$ (see Fig. 11). In Fig. 12 we compare
the likelihood for $\Omega _{m0}$ for the uncoupled case, $\delta _{0}=0$
and for the general case. Again the result is that the likelihood
for $\Omega _{m0}$ widens considerably. Now practically any value
from $\Omega _{m0}=0$ to $\Omega _{m0}=0.7$ is acceptable, with
a broad peak around 0.25, while in the uncoupled case $\Omega _{m0}$
peaks rather tightly around 0.4 (remember that we are marginalizing
over all values of $w_{\phi }$); if we restrict to $w_{\phi }>-1$
then we have roughly $0.1<\Omega _{m0}<0.3$ at 95\%. 

\begin{center}%
\begin{figure}
\begin{center}\includegraphics[  clip,
  scale=0.5]{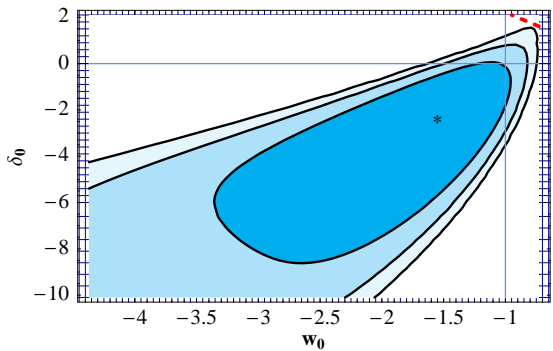}\end{center}

\caption{Likelihood contours at 68\%, 95\% and 99\% c.l., inside to outside,
marginalizing over $\Omega _{\phi }=0.7\pm 0.05$ (Gaussian prior).
In this and the following two-dimensional plots the star marks the
best fit; the horizontal line indicates the decoupled models, the
vertical line separates the phantom (on the left) from the non-phantom
models (on the right); the $\Lambda $CDM model is at the crossing
point of the two lines. Below the dashed line $\xi >0$ (fixing $\Omega _{\phi }=0.7$).
On the axes, the grid we used for the computation.}
\end{figure}
\end{center}

\begin{figure}
\begin{center}\includegraphics[  clip,
  scale=0.5]{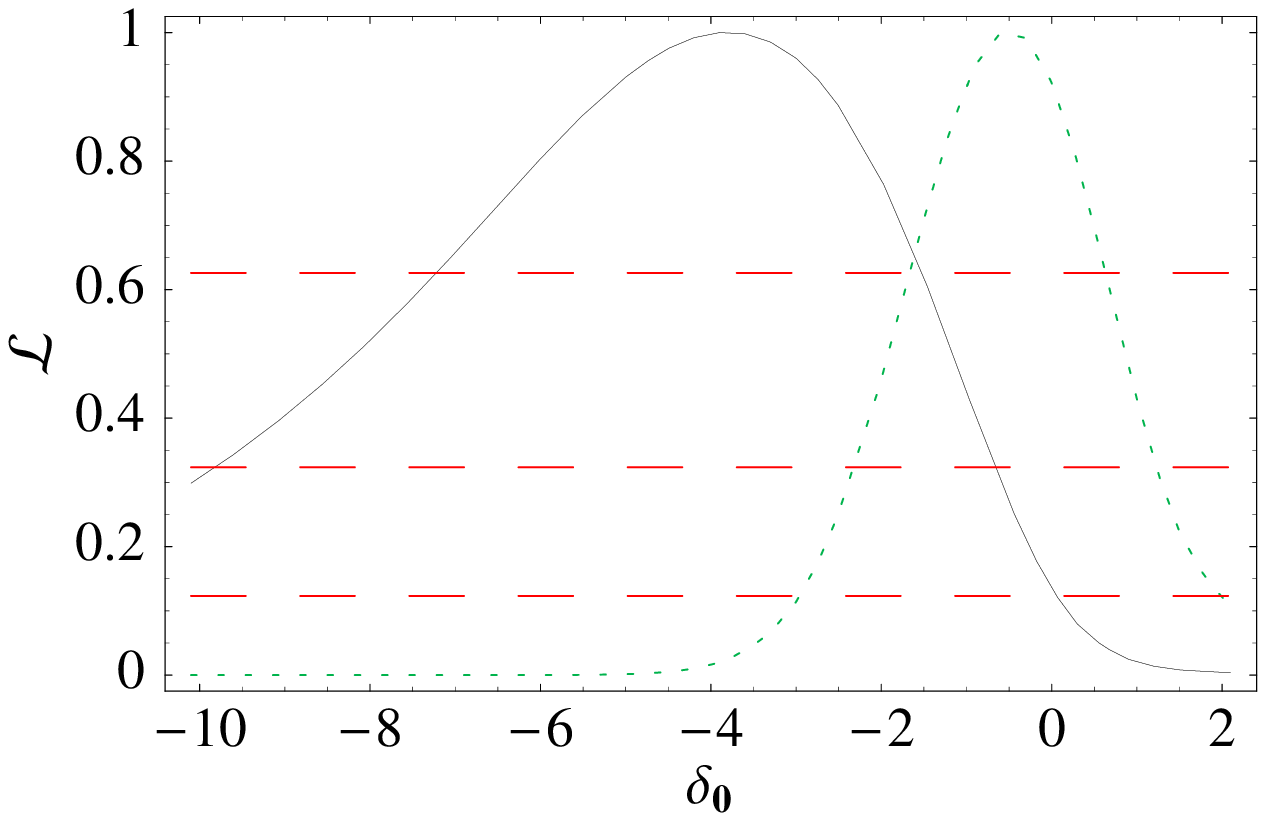}\end{center}

\caption{Full line: likelihood for $\delta _{0}$ (marginalized over $\Omega _{\phi }$
and $w_{\phi }$). The horizontal dashed lines give the 68\%, 95\%
and 99\% c.l., top to bottom. Dot-dashed line: likelihood for $\delta _{0}$
with the prior $w_{\phi }>-1$.}
\end{figure}

\begin{figure}
\begin{center}\includegraphics[  clip,
  scale=0.5]{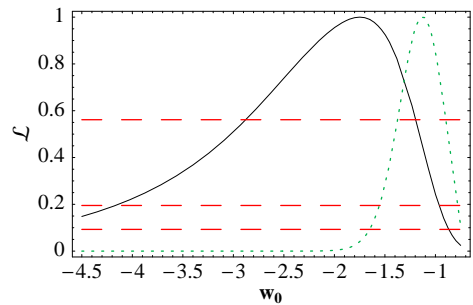}\end{center}

\caption{Full line: likelihood for $w_{\phi }$ (marginalized over $\Omega _{\phi }$
and $\delta _{0}$). The horizontal dashed lines give the 68\%, 95\%
and 99\% c.l., top to bottom. Dotted line: fixing $\delta _{0}=0$.}
\end{figure}
\begin{figure}
\begin{center}\includegraphics[  clip,
  scale=0.4]{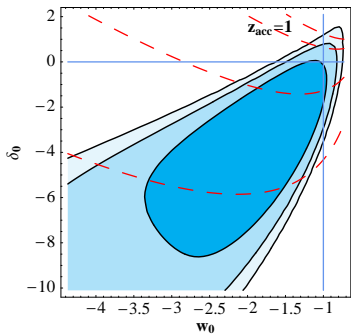}\end{center}

\caption{As in Fig. 6, with overimposed the lines of equal acceleration $z_{acc}=0.2,0.4,1,2$
(for $\Omega _{\phi }=0.7$), bottom to top.}
\end{figure}

\begin{figure}
\begin{center}\includegraphics[  clip,
  scale=0.4]{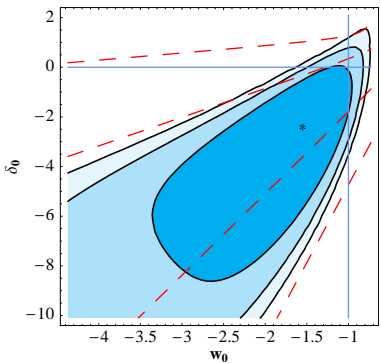}\end{center}

\caption{As in Fig. 6 with overimposed the lines of equal age $T=11,12,14,16Gyr$
(assuming $\Omega _{\phi }=0.7$, $h=0.7$) , bottom to top.}
\end{figure}

\begin{figure}
\begin{center}\includegraphics[  clip,
  scale=0.5]{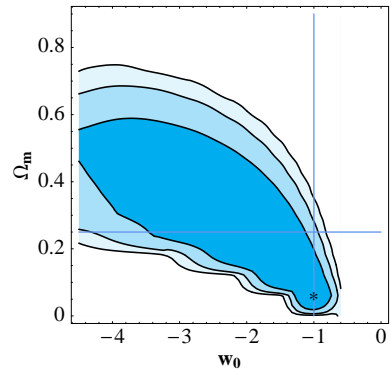}\end{center}

\caption{Likelihood for $\Omega _{m},w_{\phi }$ marginalized over $\delta _{0}$
with uniform prior in $(2,-10)$. The vertical and horizontal lines
indicate the expected values of $\Omega _{m}$ and $w_{0}$ for a
$\Lambda $CDM model.}
\end{figure}

\begin{figure}
\begin{center}\includegraphics[  clip,
  scale=0.6]{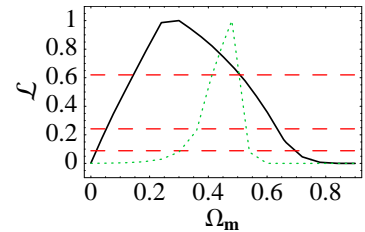}\end{center}

\caption{Likelihood for $\Omega _{m}$ marginalized over $\delta _{0},w_{\phi }$.
The dotted curve is for uncoupled dark energy ($\delta _{0}=0$).}
\end{figure}

\section{Forecasts for future esperiments}

\subsection{SNAP}

There are several project to extend the SNIa dataset both in size
and in depth. No doubt they will increase our understanding of the
dark energy problem. Here we investigate the potential to constrain
the parameters of the interacting model with a SNIa dataset that matches
the expectation from the satellite project SNAP.

We generate a random catalog of redshifts and magnitudes of 2000 supernovae
from $z=0$ to $z=1.7$, with a r.m.s. magnitude error $\Delta m=0.25$,
distributed around a $\Lambda $CDM cosmology with $\Omega _{m0}=0.3,\delta _{0}=0,w_{0}=-1$.
The redshift have been assumed uniform in the $z$-range; very likely
this is not a good approximation to what SNAP will produce but it
is rather difficult at the present stage to predict what the final
$z$-distribution will be. It is likely in fact that the statistical
significance of the larger number of expected sources at high $z$
will be at least partially reduced by additional uncertainties like
lensing effects, redshift errors, spread in the calibration curve
etc.. So we preferred to keep the forecast as simple as possible in
order not to introduce additional, and not well motivated, parameters. 

In Fig. 13 we show the likelihood marginalized over $\Omega _{m0}$.
The 68\% errors are of order 0.2 for $w_{0}$ and 0.5 for $\delta _{0}$.
Such an experiment will therefore be able to increase the precision
in $w_{0}$ in $\delta _{0}$ by a factor of five, roughly.

\begin{figure}
\begin{center}\includegraphics[  bb=0bp 470bp 598bp 843bp,
  clip,
  scale=0.6]{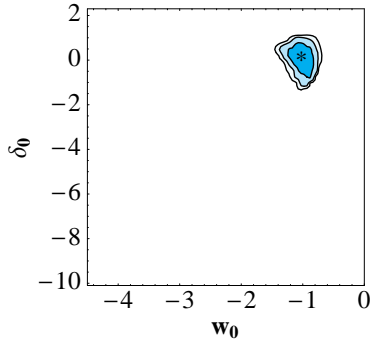}\end{center}

\caption{Likelihood contour for a SNAP-like experiment, assuming $\Lambda $CDM
($\Omega _{m}=0.3,\delta _{0}=0,w_{\phi }=-1$) as target cosmology.}
\end{figure}

\subsection{Baryon oscillations}

The main limit of the SNIa method is that even satellite experiments
like SNAP do not expect to detect a significant number of sources
beyond $z=1.7$. It has been suggested that an interesting possibility
to probe deeper the cosmic history is the reconstruction of the baryon
oscillations in the galaxy power spectrum \cite{blake}. The method
exploits the wiggles in the power spectrum induced by the acoustic
oscillations in the baryon-photon plasma before decoupling as a standard
ruler. When viewed at different $z$'s, the size of the oscillations
map into angular-diameter distances that probe the cosmic geometry
just as the luminosity distances of the SNIa. 

Here we forecast the constraints on $w_{\phi },\delta _{0}$ by baryon
oscillations assuming exactly the same experimental specification
of ref. \cite{aqg}, to which we refer for all the details. To summarize,
we evaluated the Fisher matrix of several combined datasets: five
deep surveys of 200 $\deg ^{2}$ each binned in redshift and centered
at $z=0.7,0.9,1.1,1.3,3$, plus a survey similar to the Sloan Digital
Sky Survey (SDSS) in the range $z=0-0.3$, plus a cosmic microwave
(CMB) experiment similar to the expected performance of the Planck
satellite. The five deep surveys are assumed either spectroscopic
(negligible error on $z$) or photometric (absolute error $\Delta z=0.04$).
The reference cosmology is again $\Lambda $CDM as above. With respect
to the method of \cite{aqg} we marginalize over the growth function.

The results are shown in Fig. 14 for various combinations. The label
$z=1$ denotes the combination of the four surveys at $z=0.7-1.3$;
$z=3$ denotes the furthest survey. In all cases we include the Fisher
matrices of SDSS and CMB. As expected, the deepest data are the most
powerful: the dipendence of the angular diameter distance $D$ on
$\delta _{0}$ (i.e. $\partial \log D/\partial \log \delta _{0}$)
increases by more than a factor of 2 from $z=1$ to $z=3$. Consider
first the spectroscopic case (continuous curves): including the survey
at $z=3$ pushes down the 68\% c.l. errors on $w_{\phi }$ to roughly
0.36 (not far from SNAP forecasts) and to 0.1 the error on $\delta _{0}$,
quite better than the SNAP forecasts. Adding the surveys at $z\approx 1$
 almost halves the error on $w_{\phi }$ ($\Delta w_{0}\approx 0.22$)
while is not particular effective versus the error on $\delta _{0}$.
On the other hand, the photometric surveys give constraints $\Delta w_{\phi }\approx 0.6$,
$\Delta \delta _{0}\approx 0.5$. Overall, we conclude that the baryon
oscillation method at $z\approx 3$ improves upon the SNAP experiment
for as concerns $\delta _{0}$, while has a similar efficiency in
constraining $w_{\phi }$. In Table I we list all present and future
contraints (at 68\%) we derived in this paper, bearing in mind that
the future experiments have been tested against a $\Lambda $CDM target
only: a different target cosmology implies in general different errors.

\vspace{.2in}

\begin{tabular}{|c||c||c|}
\hline 
Method&
$\Delta w_{\phi }$&
$\Delta \delta _{0}$\\
\hline
\hline 
Riess et al. (2004) SNIa&
$\approx 1$&
$\approx 2.5$\\
\hline
\hline 
Oscillations:$z=1$+$z=3$, spectr.&
0.22&
0.09\\
\hline
\hline 
Oscillations:$z=1$+$z=3$, phot.&
0.57&
0.54\\
\hline
\hline 
Oscillations:$z=1$, spectr.&
0.51&
0.81\\
\hline
\hline 
Oscillations:$z=1$, phot&
0.78&
2.52\\
\hline
\hline 
Oscillations:$z=3$, spectr.&
0.36&
0.10\\
\hline
\hline 
Oscillations:$z=3$, phot.&
0.64&
0.66\\
\hline
\hline 
SNAP&
$\approx 0.2$&
$\approx 0.5$\\
\hline
\multicolumn{3}{c}{Table I}\\
\end{tabular}

\begin{figure}
\begin{center}\includegraphics[  bb=0bp 40bp 598bp 843bp,
  clip,
  scale=0.6]{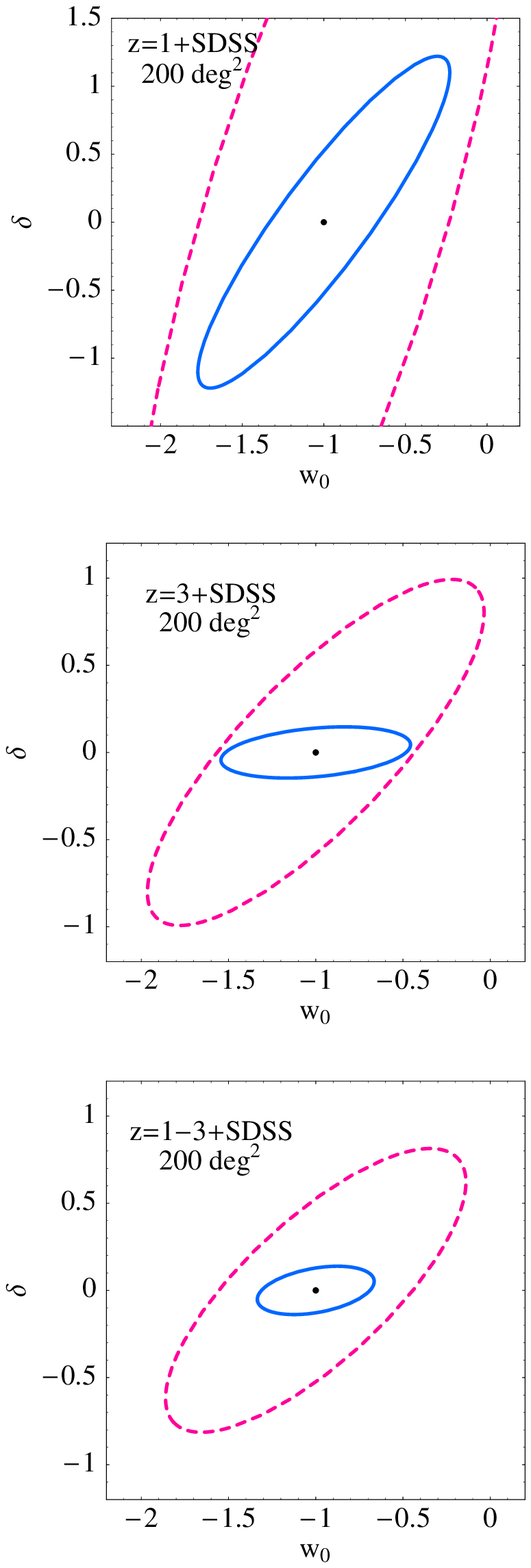}\end{center}

\caption{Likelihood contours at 68\% estimated from a Fisher matrix analysis
of baryon oscillations of several combined surveys. In all plots the
inner continuous curve is for spectroscopic surveys, the outer dashed
curve for photometric surveys with redshift error 0.04. Top: four
survey at $z\approx 1$; middle: survey at $z\approx 3$; bottom:
combined surveys at $z\approx 1$ and $z\approx 3$. In all cases,
the SDSS survey and the Fisher matrix for a CMB experiment similar
to Planck has been included.}
\end{figure}

\section{Conclusions}

This paper studied the behavior of a model with three components,
baryons, dark matter and dark energy. Contrary to most similar analyses
available in literature we included a general phenomenological coupling
between dark matter and dark energy specified by a simple but rather
general scaling relation. By this extension of dark energy models
we have been able to put bounds on the present interaction between
the two components or, equivalently, to the rate of change of the
dark matter particle mass. Moreover, we can appreciate the extent
to which the constraints on such fundamental quantitites as $\Omega _{m}$
and $w_{\phi }$ depend on the assumptions concerning the dark energy
interactions. 

We obtained several results that we summarize here.

\begin{enumerate}
\item We find the constraints on the coupling constant $\delta _{0}$, which
can be interpreted as the rate of change of dark matter mass per Hubble
time. We obtain $-10<\delta _{0}<-1$ at 95\% c.l. and $\delta _{0}<0$
at 99\%. 
\item We find $-4.2<w_{\phi }<-1$ (95\% c.l.). Together with the bounds
on $\delta _{0}$ we conclude that the current SN data set favours
negatively coupled phantom dark energy. In general, we find that the
equation of state correlates with the coupling: positive coupling
implies $w_{\phi }<-1$, negative coupling implies phantom energy.
Quite remarkably, imposing zero coupling peaks the cosmological constant
as preferred value of $w_{\phi }$.
\item Marginalizing over the coupling we derive new bounds on $\Omega _{m0}$.
In particular, we find $-4<w_{\phi }<-1$ and $0.05<\Omega _{m0}<0.65$
(95\%) . The dark matter density allowed region widens considerably
when a nonzero coupling is introduced in the model.
\item We find as best fit a value for the beginning of acceleration $z_{acc}=0.3$
but we cannot exclude earlier acceleration ($z>1$) at more than 95\%
c.l..
\item We find that the addition of even a small fraction of uncoupled matter,
e.g. baryons, modifies profoundly the asymptotic behavior of the model.
In particolar, for $\xi <0$ we find models in which the epoch of
dark energy dominance (and acceleration) is only temporary. Notice
that this occurs for $w_{\phi }$ negative and constant.
\item The ratio of baryons-to-dark matter varies in our model. Depending
on $\xi $ it may increse or decrease with time. This may offer new
methods of constraining a preferential coupling to dark matter.
\item The big rip that in uncoupled models occurs for $w_{\phi }<-1$ can
be prevented if $\delta _{0}>3|w_{\phi }|\Omega _{\phi ,0}$; however,
these values are unlikely at more than 99\% c.l.
\end{enumerate}
Future experiments will constrain the equation of state and the coupling
to a much better precision. We find that an experiment with the specification
of SNAP might reduce the errors on $\delta _{0},w_{\phi }$ by a factor
of five roughly. The method of the baryon oscillations could reduce
the error on $\delta _{0}$ by a factor of 25 roughly. Needless to
say, these forecasts depend on the exact experimental setting; however,
they give a feeling of the expected precision on the dark energy parameters
that can be reached a few years from now. Whether the final outcome
will show any trace of coupling (or, for that matter, any trace of
deviation from a pure cosmological constant) is one of the most exciting
question to ask.

\begin{acknowledgments}
We wish to thank M. Gasperini, C. Quercellini, F. Piazza for useful
discussions on topics directly related to this work.
\end{acknowledgments}


\begin{thebibliography}{10}
\bibitem{wet88}C. Wetterich Nucl. Phys. B., 302, 668 (1988)
\bibitem{wet90}C. Wetterich, Astronomy and Astrophys. \textbf{}301, 321 (1995). 
\bibitem{dam}T. Damour, G. W. Gibbons and C. Gundlach, Phys. Rev. Lett., 64, 123,
(1990) 
\bibitem{cq}L. Amendola Phys. Rev. D62, 043511 (2000)
\bibitem{chim}Chen X. \& Kamionkowsky M. Phys. Rev. D60, 104036 (1999); Uzan J.P.,
Phys. Rev. D59, 123510 (1999); Baccigalupi C., Perrotta F. \& Matarrese
S., Phys. Rev. D61, 023507 (2000); Chiba T., Phys. Rev. D60, 083508
(1999); Billyard A.P. \& Coley A.A., Phys. Rev. D61, 083503 (2000);
O. Bertolami and P. J. Martins, Phys. Rev. D 61, 064007 (2000); Faraoni
V., Phys. Rev D62 (2000) 023504; L. P. Chimento, A. S. Jakubi \& D.
Pavon, Phys. Rev. D62, 063508 (2000), astro-ph/0005070; A. B. Batista,
J. C. Fabris and R. de Sa Ribeiro, Gen. Rel. Grav. 33, 1237 (2001);
R. Bean and J. Magueijo, Phys.Lett. B517 (2001) 177 ; A.A. Sen \&
S. Sen MPLA, 16, 1303 (2001), gr-qc/0103098; W. Zimdahl, D. Pavon
\& L. P. Chimento, Phys.Lett. B521 (2001) 133, astro-ph/0105479; Chiba
T., Phys. Rev. D64 103503 (2001) astro-ph/0106550; Esposito-Farese
G. \& D. Polarsky, Phys. Rev. D63, 063504 (2001) gr-qc/0009034
\bibitem{post2002}A. Bonanno \& M. Reuter (2002) Phys. Lett. B 527, 9; P. Teerikorpi,
A. Gromov, Yu. Baryshev A\&A 407, L9 (2003) astro-ph/0209458; M. Hoffman,
astro-ph/0307350; D. Comelli, M. Pietroni and A. Riotto, Phys. Lett.
B571, 115 (2003); U. Franca \& R. Rosenfeld, Phys. Rev. D69 (2004)
063517, astro-ph/0308149; M. Axenides and K. Dimopoulos, JCAP 0407,
(2004) \textcolor{black}{hep-th}/0401238; T. Biswas \& A. Mazumdar,
hep-th/0408026
\bibitem{carroll}S.M. Carroll, Phys. Rev. Lett. 81, 3067 (1998); D. Mota \& J. Barrow,
Phys.Lett.B581:141,2004 e-Print Archive: astro-ph/0306047
\bibitem{horvat}R. Horvat, JCAP 0208:031 (2002) astro-ph/0007168
\bibitem{li}Li M., Wang X., Feng B. \& Zhang X., Phys. Rev. D65, 103511 (2002),
astro-ph/0112069 
\bibitem{gf}B.A. Gradwohl \& Frieman, J. A. 1992 ApJ, 398, 407; 
\bibitem{aq}L. Amendola\textbf{,} C. Quercellini, Phys.Rev.D68:023514, 2003 ,
astro-ph/0303228 
\bibitem{riess2004}A. G. Riess et al., ApJ, 607 (2004) 665-687, astro-ph/0402512
\bibitem{Dalal}N. Dalal et al., Phys. Rev. Lett., 87, 141302 (2001) 
\bibitem{agp}L. Amendola, M. Gasperini \& F. Piazza, JCAP 09 (2004) 007, astro-ph/0407573 
\bibitem{hagi}K. Hagiwara et al., Phys. Rev. D66 010001-1 (2002), available at pdg.lbl.gov 
\bibitem{bck}B. Bassett, P. S. Corasaniti \& M. Kunz, ApJ, 617 (2004) L1-L4, astro-ph/0407364 
\bibitem{cald}R.R. Caldwell, Phys. Lett. B545, 23 (2002)
\bibitem{barfrac}S.W. Allen et. al., Mon.Not.Roy.Astron.Soc. 353 (2004) 457, astro-ph/0405340
\bibitem{bias}L. Amendola and D. Tocchini-Valentini, astro-ph/0111535, Phys. Rev.
D66, 043528 (2002) 
\bibitem{amprl}L. Amendola, hep-th/0409224, Phys. Rev. Lett. 93 (2004) 181102
\bibitem{blake}Eisenstein D.L , Hu W. \& Tegmark M., 1999 ApJ 518, 2; Blake C. \&
Glazebrook K., 2003, ApJ 594, 665 (2003), astro-ph/0301632; Seo H.J.
\& Eisenstein D., 2003, Ap.J. 598, 720 (2003), astro-ph/0307460; Linder,
E. V. 2003, Phys. Rev. D 68, 083504 (2003); Hu W. \& Haiman Z., 2003
Phys. Rev. D 68, 063004 (2004), astro-ph/0306053; Cooray A, Huterer
D. \& Baumann D., Phys. Rev. D69, 027301 (2004)
\bibitem{aqg}L. Amendola, C. Quercellini, E. Giallongo, astro-ph/0404455, MNRAS
357, 429 (2005)\end{thebibliography}
\end{document}